\documentclass[referee]{raa}            

\usepackage{graphicx,times}             
\usepackage{natbib}

\begin{document}

   \title{Testing Lorentz violation with binary pulsars: constraints on standard model extension\,$^*$
\footnotetext{$*$ Supported by the National Natural Science Foundation of China.}
}

   \volnopage{Vol.0 (200x) No.0, 000--000}      
   \setcounter{page}{1}          

   \author{Yi Xie
      \inst{1,2}
   }

   \institute{Department of Astronomy, Nanjing University, Nanjing 210093, China; {\it yixie@nju.edu.cn}\\
	\and
	Key Laboratory of Modern Astronomy and Astrophysics, Nanjing University, Ministry of Education, Nanjing 210093, China
   }

   \date{Received~~ month day; accepted~~~~month day}

\abstract{Under the standard model extension (SME) framework, Lorentz invariance is tested in five binary pulsars: PSR J0737-3039, PSR B1534+12, PSR J1756-2251, PSR B1913+16 and PSR B2127+11C. By analyzing the advance of periastron, we obtain the constraints on a dimensionless combination of SME parameters that is sensitive to timing observations. The results imply no evidence for the break of Lorentz invariance at $10^{-10}$ level, one order of magnitude larger than previous estimation.
}

   \authorrunning{Y. Xie}            
   \titlerunning{Testing Lorentz violation with binary pulsars: constraints on SME}  

   \maketitle

%
%

\section{Introduction}
\label{sect:intro}

Unification of general relativity (GR) and quantum mechanics is a grand challenge in the fundamental physics. Some candidates of a self-consistent quantum theory of gravity emerge from tiny violations of Lorentz symmetry \citep{Kostelecky2005,Mattingly2005}. To describe observable effects of the violations, effective field theories could be a theoretical framework for tests.

The standard model extension (SME) is one of those effective theories. It includes the Lagrange densities for GR and the standard model for particle physics and allows possible breaking of Lorentz symmetry \citep{Bailey2006}. The SME parameters $\bar{s}^{\mu\nu}$ control the leading signals of Lorentz violation in the gravitational experiments in the case of the pure-gravity sector of the minimal SME. By analyzing archival lunar laser ranging data, \citet{Battat2007} constrain these dimensionless parameters at the range from $10^{-11}$ to $10^{-6}$, which means no evidence for Lorentz violation at the same level.

However, tighter constraints on $\bar{s}^{\mu\nu}$ would be hard to obtain in the solar system because the gravitational field is weak there. Thus, for this purpose, binary pulsars provide a good opportunity. Because of their stronger gravitational fields, for example the relativistic periastron advance in the double pulsars could exceed the corresponding value for Mercury by a factor of $\sim 10^5$, these systems are taken as an ideal and clean test-bed for testing GR, alternative relativistic theories of gravity and modified gravity, such as the works by \citet{Bell1996}, \citet{Damour1996}, \citet{Kramer2006}, \citet{Deng2009} and \citet{Deng2011}.

Motivated by this advantage of binary pulsars, we will try to test Lorentz invariance under the SME framework with five binary pulsars: PSR J0737-3039, PSR B1534+12, PSR J1756-2251, PSR B1913+16 and PSR B2127+11C. In Sec. \ref{sec:model}, the orbital dynamics of double pulsars in the SME will be briefed. Observational data will be used to constrain the SME parameters in Sec. \ref{sec:obs}. The conclusions and discussions will be presented in Sec. \ref{sec:con}.

\section{Orbital dynamics of double pulsars in SME}
\label{sec:model}

When the pure-gravity sector of the minimal SME is considered, it will cause secular evolutions of the orbits of double pulsars. Since timing observations of double pulsars could obtain its value very precisely, the periastron advance plays a much more important role in constraining $\bar{s}^{\mu\nu}$ and, with widely used notations in celestial mechanics, it reads \citep{Bailey2006}
\begin{eqnarray}
  \label{domegadtsme}
  \bigg<\frac{d\omega}{dt}\bigg>\bigg|_{\mathrm{SME}} & = & -\frac{n}{\tan i (1-e^2)^{1/2}}\bigg[\frac{\varepsilon}{e^2}\bar{s}_{kP}\sin\omega+\frac{(e^2-\varepsilon)}{e^2}\bar{s}_{kQ}\cos\omega-\frac{\delta m}{M}\frac{2na\varepsilon}{e}\bar{s}_k\cos\omega\bigg]\nonumber\\
  & & -n\bigg[\frac{(e^2-2\varepsilon)}{2e^4}(\bar{s}_{PP}-\bar{s}_{QQ})+\frac{\delta m}{M}\frac{2na(e^2-\varepsilon)}{e^3(1-e^2)^{1/2}}\bar{s}_Q\bigg],
\end{eqnarray}
where $M = m_1+m_2$, $\delta m = m_2-m_1$ ($m_2>m_1$) and $\varepsilon  =  1-(1-e^2)^{1/2}$. In this expression, the coefficients $\bar{s}_{\cdot}$ and $\bar{s}_{\cdot\cdot}$ for Lorentz violation with subscripts $P$, $Q$ and $k$ are projections of $\bar{s}^{\mu\nu}$ along the unit vectors $\bm{P}$, $\bm{Q}$ and $\bm{k}$. The unit vector $\bm{k}$ is perpendicular to the orbital plane of the binary pulsars, $\bm{P}$ points from the focus to the periastron, and $\bm{Q}=\bm{k}\times\bm{P}$. By definitions \citep{Bailey2006}, $\bar{s}_k\equiv \bar{s}^{0j}k^j$, $\bar{s}_Q \equiv \bar{s}^{0j}Q^j$, $\bar{s}_{kP}\equiv \bar{s}^{ij}k^iP^j$, $\bar{s}_{kQ}\equiv \bar{s}^{ij}k^iQ^j$, $\bar{s}_{PP}\equiv \bar{s}^{ij}P^iP^j$ and $\bar{s}_{QQ}\equiv \bar{s}^{ij}Q^iQ^j$. However, according to Eq. (\ref{domegadtsme}) , it is easy to see that the measurement of $\dot{\omega}$ is sensitive to a combination of $\bar{s}^{\mu\nu}$ instead of its individual components. \citet{Bailey2006} define the combination as
\begin{eqnarray}
  \label{}
  \bar{s}_{\omega} & \equiv & \bar{s}_{kP}\sin\omega+(1-e^2)^{1/2}\bar{s}_{kQ}\cos\omega-\frac{\delta m}{M}2nae\bar{s}_k\cos\omega\nonumber\\
  & &  +\tan i\frac{(1-e^2)^{1/2}(e^2-2\varepsilon)}{2e^2\varepsilon}(\bar{s}_{PP}-\bar{s}_{QQ})+\frac{m}{M}2na\tan i \frac{(e^2-\varepsilon)}{e\varepsilon}\bar{s}_Q,
\end{eqnarray}
and crudely estimate its value at the level of $10^{-11}$.

Together with the contribution from GR, the total secular periastron advance of a double pulsars system is
\begin{eqnarray}
  \label{dodttot}
  \dot{\omega} & = & 3\bigg(\frac{P_b}{2\pi}\bigg)^{-5/3}\bigg(\frac{GM}{c^3}\bigg)^{2/3}(1-e^2)^{-1}  -\frac{n\varepsilon}{\tan i (1-e^2)^{1/2}e^2}\bar{s}_{\omega}\nonumber\\
  & = &  3\bigg(\frac{P_b}{2\pi}\bigg)^{-5/3}T_{\sun}^{2/3}\bigg(\frac{M}{M_{\sun}}\bigg)^{2/3}(1-e^2)^{-1}  -\frac{2\pi\varepsilon s}{P_b (1-e^2)^{1/2}e^2 (1-s^2)^{1/2}}\bar{s}_{\omega},
\end{eqnarray}
where $T_{\sun}\equiv GM_{\sun}/c^3=4.925490947$ $\mu$s and
\begin{equation}
  \label{eqns}
  s=x\bigg(\frac{P_b}{2\pi}\bigg)^{-2/3}T_{\sun}^{-1/3}M^{2/3}m_2^{-1}.
\end{equation}
The quantity $x$ in Eq. (\ref{eqns}) is the projected semi-major axis, which is usually given by the timing observations, while, in some cases, $s$ could be measured directly so that there is no necessity to evaluate it from this equation. In this work, Eq. (\ref{dodttot}) will be taken to find the constraints on $\bar{s}_{\omega}$ with timing measurements of double pulsars.

\section{Observational constraints}
\label{sec:obs}

Long-term timing observations can determine the geometrical and physical parameters of binary pulsars very well. Among them,  PSR J0737-3039 \citep{Kramer2006}, PSR B1534+12 \citep{Staris2002}, PSR J1756-2251 \citep{Faulkner2005}, PSR B1913+16 \citep{Weisberg2010} and PSR B2127+11C \citep{Jacoby2006} are good samples for gravitational tests. Some of their timing parameters are listed in the Table \ref{Tab:timingpm}. In terms of the estimated uncertainties given in parentheses after $\dot{\omega}$, the data pool is divided into two groups: Group I, all the double pulsars are taken; and Group II, including PSR B1913+16, PSR B1534+12 and PSR B2127+11C, which have the smallest uncertainties.

By weighted least square method, the parameter $\bar{s}_{\omega}$ is estimated (see Table \ref{Tab:somega}). The estimation made by Group I is $\bar{s}_{\omega}=(-1.24\pm0.54)\times 10^{-10}$ and Group II gives $\bar{s}_{\omega}=(-1.42\pm0.75)\times 10^{-10}$. For comparison, \citet{Bailey2006} propose the attainable experimental sensitivity of $\bar{s}_{\omega}$ is $10^{-11}$, which is 10 times less than the results we obtain.

\begin{table}
\begin{center}
\caption[]{ Timing Parameters of the Double Pulsars.}
\label{Tab:timingpm}

\begin{tabular}{lllllll}
  \hline\noalign{\smallskip}
  PSR &  $P_b$(d)  &  $M$ ($M_{\sun}$)  &  $e$  &  $s$  &  $\dot{\omega}$ ($\degr$ yr$^{-1}$)  &  Reference \\
  \hline\noalign{\smallskip}
  J0737-3039 &  0.10225156248  &  2.58708  &  0.0877775  &  0.99974  &  16.89947(68)  &    \citet{Kramer2006}\\ 
  B1534+12   &  0.420737299122 &  2.678428 &  0.2736775  &  0.975    &  1.755789(9)  &    \citet{Staris2002}\\
  J1756-2251 &  0.319633898    &  2.574    &  0.180567   &  0.961${}^a$  &  2.585(2)  &    \citet{Faulkner2005}\\
  B1913+16   &  0.322997448911 &  2.828378 &  0.6171334  &  0.733650${}^a$ & 4.226598(5)  &  \citet{Weisberg2010}  \\
  B2127+11C  &  0.33528204828  &  2.71279  & 0.681395  &  0.76762${}^a$ & 4.4644(1)  &  \citet{Jacoby2006}\\
  \noalign{\smallskip}\hline
\end{tabular}
\end{center}
${}^a$Derived value according to Eq. (\ref{eqns}).
\end{table}

\begin{table}
\begin{center}
	\caption[]{ Values of $\bar{s}_{\omega}$.}
\label{Tab:somega}

\begin{tabular}{cccc}
  \hline\noalign{\smallskip}
  &  Group I  &  Group II  &  Predicted sensitivity\\
  &           &            &  \citep{Bailey2006} \\
  \hline\noalign{\smallskip}
  $\bar{s}_{\omega}$ & $(-1.24\pm0.54)\times 10^{-10}$  & $(-1.42\pm0.75)\times 10^{-10}$  &  $10^{-11}$ \\ 
  \noalign{\smallskip}\hline
\end{tabular}
\end{center}
\end{table}

\section{Conclusions and Discussion}
\label{sec:con}

In this work, we test Lorentz violation with five binary pulsars under the framework of standard model extension. It finds that $\bar{s}_{\omega}$, which is a dimensionless combination of SME parameters, is at the order of $10^{-10}$, whether all five systems are taken or top three systems with the smallest estimated uncertainties of periastron advances are used. This value, one order of magnitude greater than the estimation by \citet{Bailey2006}, implies no evidence for the break of Lorentz invariance at $10^{-10}$ level.

Nevertheless, as mentioned by \citet{Bailey2006}, the secular evolution of the eccentricity of the double pulsars should be included in the analysis. Its contribution is \citep{Bailey2006}
\begin{equation}
  \label{}
  \bigg<\frac{de}{dt}\bigg> = \frac{1}{e^3}n(1-e^2)^{1/2}(e^2-2\varepsilon)\bar{s}_e,
\end{equation}
where
\begin{equation}
  \label{}
  \bar{s}_e = \bar{s}_{PQ}-\frac{\delta m}{M}\frac{2nae\varepsilon}{e^2-2\varepsilon}\bar{s}_P.
\end{equation}
$\bar{s}_e$ is a combination of coefficients in $\bar{s}^{\mu\nu}$ and sensitive to observations. However, there is lacking of timing observations on double pulsars running for a long enough time so that rare observations could show the secular change of $e$. Even though a few numbers could be derived from data, the uncertainties of them are quite larger than those of periastron advances.  Timing observations usually could set the upper bounds only, such as $|\dot{e}|<1.9\times10^{-14}$ s${}^{-1}$ for PSR B1913+16 \citep{Taylor1989} and $|\dot{e}|<3\times10^{-15}$ s${}^{-1}$ for PSR B1534+12 \citep{Staris2002}. Hence, we suppose that, at least in current stage, the constraints made by $\dot{e}$ might be looser and the resulting upper bound is $|\bar{s}_e|<3\times 10^{-10}$. Although it is consistent with the values of $\bar{s}_{\omega}$ we obtain, the exact value of $\bar{s}_e$ remains unknown. Therefore, unless timing observations could provide much more definitive results about $\dot{e}$, the secular changes of eccentricity would not impose a tight constraint on $\bar{s}^{\mu\nu}$ or combinations of $\bar{s}^{\mu\nu}$.

Another issue for future work is to constrain the components of $\bar{s}^{\mu\nu}$ directly with double pulsars. However, the choice of reference frame affects the values of these components so that a certain reference frame must be specified first and the projections of $\bar{s}^{\mu\nu}$ will be along its standard unit basis vectors. For example, for comparing the constraints due to double pulsars and lunar laser ranging, $\bar{s}^{\mu\nu}$ has to be projected along the same triad of vectors. It means the unit vectors $\bm{P}$, $\bm{Q}$ and $\bm{k}$ (see Sec.\ref{sec:model}) have to be decomposed in terms of these vectors, which requires the geometrical information of the orbit of the double pulsars, such as the orbital elements $\Omega$ and $\omega$. Unfortunately, timing observations are not sensitive to those two elements. This makes the components of $\bar{s}^{\mu\nu}$ hard to access directly for now and demonstrates the advantages and availability of $\bar{s}_{\omega}$.

\begin{acknowledgements}
	This work is funded by the National Natural Science Foundation of China (NSFC) under Nos. 10973009 and 11103010, the Fundamental Research Program of Jiangsu Province of China under No. BK2011553, the Research Fund for the Doctoral Program of Higher Education of China under No. 20110091120003 and the Fundamental Research Funds for the Central Universities under No. 1107020116.
\end{acknowledgements}

\bibliographystyle{raa.bst}
\bibliography{gravity.bib}

\end{document}